\documentstyle[12pt]{article}
\def\u{{\mbox{\boldmath$u$}}}
\def\x{{\mbox{\boldmath$x$}}}

\def\f{{\mbox{\boldmath$f$}}}

\def\P{{\mbox{\boldmath$P$}}}
\def\k{{\mbox{\boldmath$k$}}}

\def\nab{{\mbox{\boldmath$\nabla$}}}
\def\bfpi{{\mbox{\boldmath$\pi$}}}

\def\ch{{\cal H}}
\def\dzm{{\delta\zeta_m}}

\def\begineq{\begin{equation}}
\def\endeq{\end{equation}}

\begin{document}
\title{Finite size corrections to scaling in high Reynolds number turbulence}
\author{Siegfried Grossmann $^1$, Detlef Lohse $^{2}$,\\
Victor L'vov $^{3}$, and Itamar Procaccia $^{4}$}
\maketitle

\bigskip

\begin{tabular}{ll}

$^1$ & Fachbereich Physik, Philipps-Universit\"at,\\
& Renthof 6, D-35032 Marburg, Germany \\\\
$^2$ & The James Franck Institute, The University of Chicago,\\
& 5640 South Ellis Avenue, Chicago, IL 60637, USA\\\\
$^3$ & Dept. of Physics of Complex Systems, the Weizmann Institute
of Science, \\
& Rehovot 76100, Israel \\\\
$^4$ & Dept. of Chemical Physics, the Weizmann Institute
of Science, \\
& Rehovot 76100, Israel \\\\
\end{tabular}

\date{}
\maketitle
\bigskip
\bigskip

We study analytically  and numerically the corrections to
scaling in turbulence which arise
due to the finite ratio of the outer scale $L$ of turbulence to the
viscous scale $\eta$, i.e., they are
due to finite size effects as anisotropic forcing or boundary
conditions at large scales. We find that the deviations $\dzm$
from the classical Kolmogorov scaling $\zeta_m = m/3$ of
the velocity
moments $\langle |\u(\k)|^m\rangle \propto k^{-\zeta_m}$
decrease like
$\delta\zeta_m (Re) =c_m Re^{-3/10}$.
Our numerics employ a reduced wave vector set approximation
for which the small scale structures are not fully resolved.
Within this
approximation we do not find $Re$ independent anomalous scaling
within the inertial subrange. If anomalous  scaling
in the inertial subrange can be
verified in the large $Re$ limit, this supports the suggestion
that small scale structures should be responsible, originating
from viscosity either in the bulk
(vortex tubes or sheets) or from the boundary layers
(plumes or swirls).

\vspace{0.5cm}


\newpage
A major question in the theory of turbulent flows is whether there exists
an asymptotic scaling state at high $Re$, and whether in this state the
values of the scaling exponents
conform with the classical predictions of the Kolmogorov theory
\cite{kol41}.
There is some experimental evidence that a scaling state exists,
and that the scaling exponents deviate from the Kolmogorov
prediction (anomalous scaling). Theoretically
there is still no proof for
anomalous scaling based on the
Navier-Stokes equations. Recently it was suggested that the reason for
deviations from the Kolmogorov theory is related to the
creation of small scale structures like vorticity sheets and tubes in which
the local geometry is not 3-dimensional and isotropic \cite{pro93a}.

One could hope that numerical simulations could be used to decide
whether classical or anomalous scaling should be
expected in high $Re$ flows. If anomalous scaling were found,
simulations could distinguish the responsible physical
mechanism.
Unfortunately, in direct simulations of the Navier-Stokes equations
with current computers one can achieve only up to Taylor-Reynolds
numbers of about $Re_\lambda = 200$ \cite{she93}. The range of scales
for which power law behavior is observed at such values of $Re_\lambda$
is too small to distinguish between classical and anomalous scaling.
In addition, at low to moderate values of $Re$  one can have
corrections to scaling due to finite size effects, and it is hard to
take these into account if their expected $Re$ dependence is not known.
It is this latter issue which is the focus of this letter. We examine
these corrections to scaling analytically and numerically, and show
how to take them into account in any future experiment or
numerical simulation.

A high $Re$ flow can be constructed in a numerical simulation by using
a reduced wave vector set approximation
(Fourier-Weierstrass decomposition),
that was introduced and studied extensively
by two of us recently.
For a detailed description
and references of previous work we
refer to refs.\ \cite{gnlo92b,gnlo93e,gnlo94b}. The main idea is
to start with a regular Fourier decomposition of the velocity
field $\u (\x ,t)$ in terms of plane waves $\exp{(i\k\cdot\x)}$,
but admit only a geometrically scaling subset $K = \cup_l K_l$
of wave vectors in the Fourier sum,
\begineq
\u(\x ,t) = \sum_{k\in K} \u (\k ,t) \exp{(i\k \cdot\x)}.
\label{eq1}
\endeq
The subset of wave vectors is chosen such that
$K_0=\{k_n^{(0)}, n=1,\dots,N\}$ and
$K_l=\{2^lk_n^{(0)}, n=1,\dots,N\}$, $l=1,\dots , l_{max}$. The basic
set $K_0$ is chosen to contain wave vectors of different lengths
that interact dynamically to a good degree, and $l_{max}$ is
chosen large enough to guarantee that the amplitudes
$\u (\k_n^{(l_{max})},t)$ of the smallest scales are practically zero.
This of course depends on the viscosity $\nu$ and the Reynolds number
$Re$. The flow is driven by a deterministic, non stochastic driving
$\f (\k ,t)$ with $\k$ in $K_0$. In fact, only the seven smallest
wave vectors in $K_0$ were driven. Obviously, this driving is not
isotropic.
The $k$-range of externally forced amplitudes is denoted henceforth
as stirring subrange.

The simulations that we refer to here used a value of $N=80$ and scanned
a range of $Re$ from $10^4$ to $10^7$ \cite{gnlo94b}. It should
be stressed that in this scheme the density of wave vectors per
k interval decreases like $1/k$, whereas it increases like $k^2$
in full grid simulations. For this reason small scale structures
(as vortex tubes or sheets) cannot
be resolved very accurately. But on the other hand many more scales
than in full simulations can be examined. For the largest $Re$
our simulations comprise more than three orders of magnitude
between the wave number with maximum dissipation and the outer scale.
The scaling exponents $\zeta_m$ can be determined with high accuracy.

The scaling exponents were measured in this numerical simulation
by computing the $m^{th}$-order spectra $\langle |\u (\k ) |^m\rangle$,
and fitting the three parameter function
\begineq
\langle | \u ( \k )|^m \rangle
= c_m  k^{-\zeta_m} \exp{[-  k / k_d^{(m)}]}.
\label{eq2}
\endeq
It was found \cite{gnlo93e,gnlo94b} that a good approximation could be
obtained globally with
\begineq
k_d^{(m)}=2k_d^{(2)}/m, \qquad k_d^{(2)}=(13.5\eta)^{-1},
\label{eq3}
\endeq
with $\eta $ being the Kolmogorov length scale.
Eq.\ (3) means that for
higher order velocity
moments the crossover between
the inertial range and the viscous range  takes place at
smaller value of $k$, i.e., the inertial range is less extended.
As we want to examine the effects of the large scales, i.e., of the
stirring subrange, we eliminate viscous
effects by
fitting (2) with fixed $k_d^{(m)}$ (according to (3)) {\it only}
in the interval $[0,k_d^{(m)}]$. The resulting
scaling exponents are (very slightly) corrected
to guarantee $\zeta_3=1$, see \cite{gnlo93e,gnlo94b} for the procedure.
$\zeta_3 = 1$ is required by Kolmogorov's structure equation \cite{my75},
which strictly holds only
in an isotropic situation. But as $\zeta_3=1$ is
frequently enforced in the analysis of
experimental data, we do it here as well. We represent
the numerical results for the scaling exponents in terms of
the deviations from the classical predictions
$ \zeta_m (Re) - m/3$.
The results, which pertain now to both
inertial range and stirring range, are
shown in fig.\ 1 for four different $Re$. The deviations
are decreasing with increasing $Re$.

Qualitatively this feature can be understood from the comparison
with
the {\it local} $\dzm (k)$ for each $Re$, see ref.\ \cite{gnlo93e}.
The $\dzm (k)$ are calculated by {\it locally}
employing the fit (2) to the
spectra with $k_d^{(m)}$ fixed. The results are shown in fig.\ 2a
for $Re=1.05 \cdot 10^4$ and in fig.\ 2b for $Re= 1.4\cdot 10^7$.
Clearly, with increasing Re the inertial range
extension exceeds more and more that of the stirring range,
and thus the effect of the nonuniversal forcing diminishes.

To get quantitative information, we fitted the power law
\begineq
\dzm (Re) = c_m Re^{-\beta_m}
\label{eq4}
\endeq
to the $\dzm (Re)$ determined from the spectra.
The results of the fit (4)
for $c_m$ and $\beta_m$ are given in table 1. We find that
$\beta_m$ is close to $3/10$ for all $m$.

A good way to understand this behavior is to consider
the Euler part of the equations of motion using the representation
by the Clebsch variables. In these variables the velocity field is
written in terms of the two scalar functions $\lambda (\x ,t)$
and $\mu (\x ,t)$:
\begineq
\u(\x,t)=\lambda \nab\mu - \nab \phi,
\label{eq5}
\endeq
where the potential $\phi (\x ,t)$ is determined from the
incompressibility condition. Using these variables, the equation of
motion
\begineq
\partial_t\u(\x,t)+\u\cdot\nab\u=-\nab p(\x,t)
\label{eq6}
\endeq
reads
\begineq
\partial_t \lambda (\x ,t) = {\delta \ch \over \delta \mu (\x ,t)},
\qquad
\partial_t \mu (\x ,t) =- {\delta \ch \over \delta \lambda (\x ,t)},
\label{eq7}
\endeq
where the Hamiltonian $\ch$ is defined as
$\ch =  \int d\x |\u(\x ,t)|^2 /2.$
This formulation allows to introduce normal canonical
variables $a(\x,t)$ and $a^*(\x,t)$, cf.\ \cite{gro75},
\begineq
\sqrt{2} a(\x,t)=\lambda (\x ,t) + i \mu (\x,t).
\label{eq9}
\endeq
These canonical variables can be used to expose integrals of motion.
In particular, the ``occupation number'' $N$
and the ``momentum'' $\P$ of quasi particles,
\begineq
N=\int |a_k|^2 d\k,
\qquad
\P=\int \k |a_k|^2 d\k,
\label{eq11}
\endeq
are such integrals of motion \cite{lvo94,lvo94a}. These integrals of motion
could also be expressed as functions of the velocity field $\u (\x,t)$,
but are highly nonlocal in that representation. The main point is
that the integral $\P$ can be non zero only in an anisotropic system, and
therefore can lead us to scaling laws that capture the rate of decay
of anisotropy as a function of $\k$. To proceed in this direction we
first find by dimensional analysis that in an isotropic system the
k-component of $N$, $N_k$, depends on $k$ and the mean energy flux per
unit time per unit mass, $\epsilon$, according to
\begineq
N_k\delta (\k-\k') = \langle a_k a^*_{k'}\rangle
=C \epsilon^{1/3} k^{-13/3}\delta (\k-\k')
\label{eq12}
\endeq
with $C$ being a dimensionless constant. Of course, this result
can be translated back to the
standard Kolmogorov result for the energy
$E_k = C\epsilon^{2/3} k^{-5/3}$. Next, we realize that in an anisotropic
system this relation can change, since $N_k$ can depend now on a
dimensionless ratio of the ``momentum flux'' over the energy flux
$\epsilon$. The energy flux is defined by the equation
\begineq
\partial_t \ch_k + {\partial \epsilon_k \over \partial k}=0\,,
\label{eq13}
\endeq
which in a stationary state requires $\epsilon_k$ to be $\k$-independent,
$\epsilon_k =\epsilon$. The momentum flux $\bfpi_k$, likewise, is defined
by
\begineq
-\partial_t \P_k =
\k k^2 \partial_t N_k(t)=
{\partial \bfpi_k \over \partial k}.
\label{eq14}
\endeq
In a stationary system $\bfpi_k$ is also $k$-independent, and from eqs.\
(\ref{eq12}) and (\ref{eq14})
we see that the dimension of $\pi$ can be written as
$[\pi ]= [\epsilon^{2/3} k^{1/3}]=  length/(time)^2.$
In an anisotropic system we seek a new solution for $N_k$,
\begineq
N_k = C \epsilon^{1/3} k^{-13/3} f(\xi_k)
\label{eq16}
\endeq
where the dimensionless parameter  $\xi_k$ is proportional to the
ratio of the two fluxes $\pi$ and $\epsilon$. Since the two fluxes
$\epsilon$ and $\pi$ have different dimensionalities, the parameter
$\xi_k$ must involve $k$ and $\epsilon$ to some powers. The unique
combination that is proportional to $\pi$ is
\begineq
\xi_k = {\pi \over \epsilon^{2/3} k^{1/3}}\ .
\label{eq17}
\endeq
The function $f(\xi_k)$ is chosen such that for $\pi=0$ Eq.\
(\ref{eq16}) regains its Kolmogorov form, i.e. $f(0)=1$.
Asuming that for small $\pi$ the correction to the spectrum is
proportional to $\pi$, for small $\xi$ we choose $f(\xi)=1 +O(\xi)$.
Notice that in (\ref{eq16}) and
(\ref{eq17}) we assumed that the anisotropic correction to $N_k$ is
analytic in $\pi$. This is of course a crucial assumption, but it
can be justified by an explicit calculation, and it was shown to be
correct to all orders in perturbation theory, see \cite{lvo94}.
Denoting the anisotropic correction as $\delta N_k$, we conclude that
for small anisotropy, $\xi \ll 1$,
\begineq
\delta N_k / N_k \propto k^{-1/3}.
\label{eq18}
\endeq
Obviously, since $\bfpi_k$ is odd in $\k$, cf.\ (\ref{eq11})
or (\ref{eq14}), so is also $\delta N_k$.
In contrast, the double correlation function of the velocity field
$F(\k)$,
\begineq
F(\k)\Delta (\k - \k') =Tr { \langle \u (\k) \u^* (\k')\rangle}
\label{eq19}
\endeq
is even in $\k$.
(Note that in accordance with our discrete simulations
we define $F(\k)$ with a Kronecker $\Delta$.) Now, since $F(\k)$
is even in $\k$,  the lowest order correction due to anisotropy
must be the second order. We can conclude therefore that
\begineq
\delta F (\k) / F(\k) \propto k^{-2/3}.
\label{eq20}
\endeq
To bring this result to a form that can be tested against the simulation,
we multiply $F(\k)$ by $k$ (as numerically we of course
use discrete wave vectors)
and rewrite our result in the form
\begineq
F(\k)=
\langle | \u ( \k )|^2 \rangle
=c_2  k^{-2/3}
\left( 1+\alpha_m (kL)^{-2/3} \right)
 \exp{[-  k / k_d^{(2)}]}.
\label{eq21}
\endeq
($L$ is the outer length scale.)
The same conclusion is obtained by considering the effect of external shear
\cite{lum72}. Since it introduces another time scale $\tau (k_L)$,
spectral corrections should be
$\propto \tau (k)/\tau(k_L) \propto k^{-2/3}$.
On scales on which the shear dominates ($k\ll L^{-1}$) the second
term in (\ref{eq21}) will be even the leading contribution
and thus will change the dominating $k$-power to $k^{-4/3}$.
For a similar recent result concerning the $k^{-4/3}$-scaling law for small
$k$ in an anisotropic flow see \cite{yak94}.
The crossover from
$k^{-4/3}$ to $k^{-2/3}$ can well be observed in experiments
\cite{yak94}. Here we restrict ourselves to $k>L^{-1}$.
So the second term in (\ref{eq21}) is an anisotropy correction, which will
become smaller for increasing $k$ due to isotropization
by eddy decay.
Yet it will
render the scaling exponent k-dependent.
Denoting the local scaling exponent
$dln\langle|\u (\k)|^2\rangle /dlnk$
as $\zeta_2(k)$, we find from (\ref{eq21})
\begineq
\zeta_2 (k) = {2\over 3} + {k \over k_d^{(2)}} +
{2\over 3[1+\alpha_2^{-1} (kL)^{2/3}]}.
\label{eq22}
\endeq
As a function of $k$ this expression has a minimum which is
rather flat, and can lead to  an apparent  exponent $\zeta_2^{(app)}$.
We estimate the value of this exponent by the minimum of (\ref{eq22}) which
approximately is
\begineq
\zeta_2^{(app)} = {2\over 3} +  {10\over 9}
\left( {9\over 4  k_d^{(2)}L} \right)^{2/5}
|\alpha_2|^{3/5}.
\label{eq23}
\endeq
Here, we used the fact that $1\ll \alpha_2^{-1} (kL)^{2/3}$
near the minimum, justified a posteriori,
as $k_d^{(2)}L \gg 9\alpha_2^{3/2}/4$.
 From (\ref{eq23}),  we can predict that the deviation of $\zeta_2^{(app)}$
from $2/3$ goes down when $Re$ increases,
since $k_d^{(2)}L\propto Re^{3/4}$
\cite{my75}.
We obtain
\begineq
\delta\zeta_2^{(app)} \propto Re^{-3/10}.
\label{eq24}
\endeq
Exactly the same calculation can be repeated for all other
scaling exponents $\zeta_m$, and all of them approach their
Kolmogorov 41 values with the same scaling $\propto Re^{-3/10}$.
The result of the calculation is
\begineq
\zeta_m^{(app)} = {m\over 3} + sgn(\alpha_m) {10\over 9}
\left( {9\over 4  k_d^{(m)}L} \right)^{2/5}
|\alpha_m|^{3/5},
\label{eq25}
\endeq
\begineq
\delta\zeta_m^{(app)} =c_m Re^{-3/10}
\label{eq26}
\endeq
with
\begineq
{c_m\over c_2} = sgn({\alpha_m\over \alpha_2} )
({m\over 2})^{2/5} |{\alpha_m\over\alpha_2}|^{3/5} .
\label{eq27}
\endeq
 From eq.\ (\ref{eq27}) and table 1 we can determine the m-dependence of
$\alpha_m$. Assuming a power law dependence, we estimate
$\alpha_m \propto m^{3.8\pm 0.8}$, which is a rather strong
m-dependence.
This finding for $\alpha_m$ means that
higher order moments are more affected by corrections to scaling
than lower order ones.
Together with the fact that their power law behavior is cut off at
lower k vectors, cf.\ eq.\ (3), it means that it is harder to observe clean
power laws
in higher order moments
for an appreciate range of scales.

In summary, the results of the numerical simulations of the Navier-Stokes
equation with a reduced wave vector set approximation are in very
close correspondence with the predictions of a theory in which the deviations
from classical scaling is solely due to corrections to scaling. It
is tempting to conjecture that if a full treatment
of the Navier-Stokes
dynamics results in genuine anomalous scaling that does not
disappear in the limit of high $Re$, it has to do with the small scale
structures that are underestimated in our simulations as well
as in the analytical argument that refered to the Euler equations and
neglected viscosity and boundary layers. Such a conjecture would be in
accord with other theoretical expectations which link anomalous scaling
to the creation of vortex structures like tubes and sheets.

\vspace{1.5cm}
\noindent
{\bf Acknowledgements:}
Support by the
German-Israel-Foundation (GIF),
the Minerva Center of Nonlinear Physics,  and by the
Deutsche Forschungsgemeinschaft, SFB185,
is gratefully acknowledged.
D.L. also acknowledges support  by DOE and by
a NATO grant, attributed by the Deutsche
Akademische Austauschdienst (DAAD).

\newpage

\centerline{\bf Tables}

 \begin{table}[htp]
 \begin{center}
 \begin{tabular}{|r|r|r|r|r|r|}
 \hline
         $ m  $
       & 2
       & 4
       & 6
       & 8
       & 10
 \\
\hline
         $ \beta_m  $
       & 0.24
       & 0.29
       & 0.28
       & 0.31
       & 0.30
 \\
         $ c_m  $
       & $+0.11$
       & $-0.32$
       & $-1.30$
       & $-4.03$
       & $-7.02$
\\
 \hline
 \end{tabular}
 \end{center}
 \end{table}

\vspace{0.5cm}

\centerline{\bf Table 1} \noindent
Fitparameters to  eq.\ (4). If the m-dependence of
$c_m$ is expressed by a power law, we will get $c_m\propto
m^{2.7\pm0.5}$.
The data are compatible with $c_m\propto (m-3)^{3\pm 0.5}$,
indicating agreement with the Kolmogorov structure equation,
which forbids
corrections if $m=3$.
Here also the change of sign of the $c_m$ is immediately grasped.

\vspace{2cm}

\centerline{\bf Figures}
\begin{figure}[htb]
\caption[]{
Results from our reduced wave vector set
approximation for $|\dzm (Re)|$ for m=2,4,6,8,10, bottom to top.
The straight lines correspond to the scaling law $\dzm (Re)
\propto Re^{-3/10}$ as predicted by (\ref{eq26}).
}
\label{f_dzm}
\end{figure}

\begin{figure}[htb]
\caption[]{Scale resolved intermittency corrections $-\delta\zeta_m(k)$ for
$m=2,4,6,8,10$, bottom to top. In (a) we have
$Re= 1.05 \cdot 10^4$, in (b) it is $Re= 1.4 \cdot 10^7$.
The fit range is
$[ k/\sqrt{10}, k\sqrt{10} ]$ for all $k$.
For details see ref.\ \cite{gnlo93e,gnlo94b}.
}
\label{f_sri}
\end{figure}

\newpage


\begin{thebibliography}{10}

\bibitem{kol41}
A.~N. Kolmogorov, CR. Acad. Sci. USSR. {\bf 30},  299  (1941);
A.~M. Obukhov, Izv. Akad. Nauk SSSR, Ser. Geog. Geofiz. {\bf 13},  58  (1949);
C.~F. von Weizs\"acker, Z. Phys. {\bf 124},  614  (1948);
W. Heisenberg, Z. Phys. {\bf 124},  628  (1948);
L. Onsager, Phys. Rev. {\bf 68},  286  (1945).

\bibitem{pro93a}
I. Procaccia and P. Constantin, Phys. Rev. Lett. {\bf 70},  3416  (1993);
P. Constantin and I. Procaccia,
``Creation and dynamics of vortex tubes in 3-dimensional
turbulence'', Phys. Rev. E  (1994), in press.

\bibitem{she93}
Z.~S. She, S. Chen,  G. Doolen,  R. H. Kraichnan,  and S. A. Orszag,
Phys. Rev. Lett. {\bf 70},  3251  (1993).

\bibitem{gnlo92b}
S. Grossmann and D. Lohse, Z. Phys. B {\bf 89},  11  (1992).

\bibitem{gnlo93e}
S. Grossmann and D. Lohse,
``Scale resolved intermittency in turbulence'',
Phys. Fluids A  (1994), in press.

\bibitem{gnlo94b}
S. Grossmann and D. Lohse,
``Universality in fully developed turbulence'',
Phys. Rev. E  (1994), submitted.

\bibitem{my75}
A.~S. Monin and A.~M. Yaglom, {\em Statistical Fluid Mechanics} (The MIT Press,
  Cambridge, Massachusetts, 1975).

\bibitem{gro75}
S. Grossmann, Phys. Rev. A {\bf 11},  2165  (1975).

\bibitem{lvo94}
V.~S. L'vov and G.~A. Falkovich, Chaos, Solitons, and Fractals
(1994), in press.

\bibitem{lvo94a}
V.~S. L'vov and I. Procaccia,
``Extended universality in moderate Reynolds number flows'',
Phys. Rev. E  (1994), in press.

\bibitem{lum72}
H. Tennekes and J.~L. Lumley, {\em A first course in turbulence} (The MIT
  Press, Cambridge, Massachusetts, 1972).

\bibitem{yak94}
V. Yakhot, ``Large scale coherence and anomalous scaling of
high-order moments of velocity differences in strong turbulence'',
preprint, 1993.

\end{thebibliography}


\end{document}